\documentclass[prd,nofootinbib,twocolumn,showpacs]{revtex4}  
\usepackage[dutch,german,british,american]{babel}
\usepackage {amsmath}
\usepackage{mathrsfs,graphicx,verbatim}
\usepackage{epsfig}
\usepackage{comment}

\newcommand {\be}{\begin{eqnarray}}
\newcommand {\ee}{\end{eqnarray}}

\def\ba{{\bf a}}
\def\bx{{\bf x}}
\def\bn{{\bf n}}

\def\D{\nabla}
\def\tD{\tilde{\nabla}}
\def\U{\mathcal{U}}

\begin{document}

%\title{The Link Approach to Supersymmetry on the Lattice:\\
%$N=D=2$ SYM and an Inconsistency}
\title{A Critique of the Link Approach to Exact Lattice Supersymmetry
}
\author{Falk Bruckmann}
\affiliation{Institute for Theoretical Physics, University of Regensburg}

\author{Simon Catterall}
\affiliation{Department of Physics, Syracuse University, Syracuse NY}

\author{Mark de Kok}
\affiliation{Instituut-Lorentz for Theoretical Physics, Leiden University}

\date{\today}% It is always \today, today,

\begin{abstract}
We examine the link approach to constructing a lattice theory
of ${\cal N}=2$ super Yang Mills theory in two dimensions. The goal
of this construction is to provide a discretization of
the continuum theory which
preserves all supersymmetries at non-zero lattice
spacing. We show that this approach suffers
from an inconsistency and
argue that
a maximum of just one of the supersymmetries can be implemented
on the lattice.
\end{abstract}

\pacs{11.30.Pb, 11.15.Ha}% PACS, the Physics and Astronomy
\maketitle

%%%%%%%%%%%

\section{Introduction}

For the study of non-perturbative effects in supersymmetric field theories, 
important for e.g.\ supersymmetry breaking, it is very useful to have lattice formulations of these theories. Such formulations allow one to  
do computer simulations. From a practical
point of view the main problem that arises when one tries to
write down such lattice theories lies in the failure of the ordinary product rule of differentiation, as explained in e.g.\ \cite{Giedt}. This generically leads
to lattice actions which classically break all the supersymmetries of the
continuum theory. This in turn renders recovering a supersymmetric continuum
limit problematic -- typically the lattice action must be supplemented by
the addition of large numbers of relevant supersymmetry violating terms whose
couplings must be carefully fine tuned as the lattice spacing is reduced.

In light of this a couple of recent approaches to the problem of lattice
supersymmetry attempt to preserve a fraction of the original supersymmetry
exactly at non-zero lattice spacing \cite{Kaplan, Catterall, Sugino}.
In one approach a supersymmetric lattice action is constructed by
orbifolding a supersymmetric matrix model while the other
proceeds by finding combinations of the original supercharges which
behave like internal symmetries and hence can be transferred to
the lattice. While these two
constructions appear quite different, they have recently been shown to
be intimately related and to correspond to discretizations of {\it twisted}
formulations of the continuum supersymmetric theories \cite{Unsal}.

In \cite{Dadda} an alternative discretization of these 
twisted theories was described,
which aimed at repairing the product rule of differentiation, the Leibniz rule, by introducing a non-commutativity between the bosonic and fermionic coordinates
of twisted superspace. The claim is that lattice theories constructed
in this way
exhibit invariance under the {\it full} set of continuum supersymmetries.
However,
as explained in \cite{FalkMe}, unfortunately
this approach suffers from an inconsistency.

Another approach related to these non-commutative
lattice formulations is the link approach, which was introduced by
the same authors in \cite{Dadda3}. Because the paper 
\cite{FalkMe} never addressed the link approach, 
a certain amount of
confusion and discussion has arisen
as to whether this approach is also inconsistent or not. 
In this paper we describe this
construction and show that a similar problem does indeed arise in this
case, too.

In the following section a summary of the non-commutativity approach
and its inconsistency is given. 
The third section describes the link approach, and the fourth section
highlights an apparent inconsistency inherent also in that
construction. 
The paper ends with conclusions and some discussion.

%%%%%%%%%%%%%%%%%%%%%%%%%%%

\section{The NC Approach}

Instead of having derivative operators, on the lattice one has to deal with difference operators.
For example, for the forward and backward difference operator, acting on
functions $f$ on a lattice with coordinates $x^{\mu}$ as
\be
\Delta_{\pm\mu}f(\bx) = \pm\frac{1}{|\bn_{\mu}|}\left(f(\bx\pm \bn_{\mu}) - f(\bx) \right),
\ee
(where $\bn_{\mu}$ corresponds to the shift of one lattice spacing in the $\mu$-direction)
the following product rule holds
\begin{equation}
\begin{array}{ccl}
\Delta_{\pm\mu}\left[f_1(\bx)f_2(\bx)\right] & = & \left[\Delta_{\pm\mu}f_1(\bx)\right]f_2(\bx)  \\[8pt] 
 & &\quad +f_1(\bx\pm \bn^{\mu})\left[\Delta_{\pm\mu}f_2(\bx)\right].
\end{array}
\end{equation}
This differs from the ordinary Leibniz rule by a shift in the argument of one of the functions.

The non-commutativity (NC) approach, as introduced in \cite{Dadda}, can now be explained as follows.
For writing down supersymmetric lattice theories, the real necessity for having the Leibniz rule
lies in having this rule hold for the supersymmetry variations, not just the difference operator. 
The main idea is to introduce a non-commutativity between the bosonic and fermionic coordinates in a superspace representation, such that the combinations $\delta_A = \epsilon^AQ_A$ (no sum), where the
$Q_A$ are the supercharges and the $\epsilon^A$ Grassmannian susy parameters, obey the Leibniz rule.

Ignoring the Lorentz transformations, the general supersymmetry algebra can be written as
\be
\left\{Q_A,Q_B \right\} = f^{\mu}_{AB}\Delta_{\pm\mu}, \quad [Q_A, \Delta_{\pm\mu}] = 0.
\ee
As explained in \cite{Dadda,FalkMe} the supersymmetry variations $\delta_A$ will obey the Leibniz
rule if the following non-commutativities between the coordinates of superspace and the susy parameters are introduced
\be
[\bx,\theta^{A}] = \ba_A\theta^{A}, \quad [\bx,\epsilon^{A}] =
\ba_A\epsilon^{A}\quad \mbox{(no sum)}.
\label{eq_nc_general}
\ee
This can only be done consistently if the following equations hold:
\be
\ba_A + \ba_B = \pm \bn_{\mu} \mbox{ for } f^{\mu}_{AB} \neq 0. 
\label{relations}
\ee
The $a_A$ are called the shift parameters.

D'Adda et al.\
have found a solution for these relations for the twisted
${\cal N}=D=2$ and twisted ${\cal N}=D=4$ supersymmetry algebras.
As explained in \cite{FalkMe}, these relations 
can also be satisfied for the ${\cal N}=2$ supersymmetry algebra in one dimension, supersymmetric quantum mechanics.

Let's look in more detail at the ${\cal N}=D=2$ case. Reformulating the algebra by twisting 
\cite{Dadda%,Uchida,Witten
}, it reads
\be
\{ Q , Q_{\mu} \} = i \Delta_{+\mu}, \qquad
\{\tilde{Q}, Q_{\mu} \} = -i \epsilon_{\mu \nu} \Delta_{- \nu},
\label{discrsusy}
\ee
where only the non-vanishing anticommutators are shown.  
Introducing the supercoordinates $\theta, \tilde{\theta}, \theta^{\mu}$,
 the supercharges read
\begin{align}
Q&=\frac{\partial}{\partial \theta} +
\frac{i}{2} \theta^{\mu}\Delta_{+\mu}, \nonumber \\
 \tilde{Q}&=\frac{\partial}{\partial\tilde{\theta}} -
\frac{i}{2}\epsilon_{\mu \nu} \theta^{\mu}\Delta_{-\nu}, \label{susych}%\nonumber 
\\
Q_{\mu}&=\frac{\partial}{\partial \theta_{\mu} }+ \frac{i}{2} \theta\Delta_{+\mu}
 - \frac{i}{2} \epsilon_{\mu \nu} \tilde{\theta} \Delta_{-\nu}.\nonumber
\end{align}

\begin{figure}[t]
\hfil \includegraphics[width=35mm]{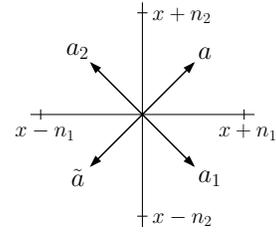}
\caption{Symmetric choice of shift parameter $a_{A}$ for twisted ${\cal N} = D = 2$.}
\label{symmetric_a}
\end{figure}

Demanding that independently $\epsilon Q$, $\tilde{\epsilon}\tilde{Q}$, $\epsilon^1Q_1$ and
$\epsilon^2Q_2$ satisfy the Leibniz rule, one is lead to the following non-commutativities
\be
[x^{\mu}, \theta] = a\theta, \, [x^{\mu}, \tilde{\theta}] = \tilde{a}\tilde{\theta}, \, [x^{\mu}, \theta^{\nu}] = a^{\nu}\theta^{\nu} \, \mbox{(no sum)},
\ee
and similarly for the $\epsilon_A$.
For consistency the shift parameters have to satisfy the following conditions:
\be
&a + a_{\mu} = n_{\mu}, \quad \tilde{a}+ a_{\mu}=-|\epsilon_{\mu \nu}| n_{\nu},& \nonumber\\[8pt]
&a+\tilde{a}+a_1+a_2=0.&
\label{consiscond}
\ee
These conditions have a symmetric solution 
\be
a= -\tilde{a} = (1/2,1/2), \quad a_1=-a_2=(1/2,-1/2), 
\label{symmsolution}
\ee
as shown in 
figure~\ref{symmetric_a}.
Working with this solution one is thus forced to introduce new lattice points at the
half integer lattice sites, effectively doubling the lattice. 
Actually, one of the parameters is left undetermined by the equations (\ref{consiscond}), leading to a linear space
of solutions.
Another interesting solution is given by
\be
a= (0,0),\, \tilde{a} = (-1,-1),\, a_1=(1,0), \,a_2=(0,1),
\label{azerosolution}
\ee
where one of the shift parameters vanishes and all others point to already
existing lattice points.
\\
\newline
However, the non-commutativity approach is plagued by an inconsistency. This inconsistency
holds in general, as explained in \cite{FalkMe}. 
Since the $\epsilon^AQ_A$ (no sum) obey the Leibniz rule, also the $\epsilon^As_A$ will obey the Leibniz rule, where the $s_A$ denote the supersymmetry transformations of the component fields. 
It thus holds that
\begin{equation}
\begin{array}{l}
\epsilon^As_A\left[f_1(\bx)f_2(\bx)\right]  \\[8pt]
= \left[\epsilon^As_A f_1(\bx)\right]f_2(\bx)
+ f_1(\bx)\left[\epsilon^As_A f_2(\bx)\right],
\end{array}
\end{equation}
for any two component fields $f_1$ and $f_2$.
Changing the order in the product, it just as well holds that
%\begin{equation}
%\begin{array}{l}
%\epsilon^As_A\left[f_1(\bx)f_2(\bx)\right]  \\[8pt]
%= \epsilon^As_A\left[(-1)^{|f_1||f_2|}f_2(\bx)f_1(\bx)\right]  \\[8pt]
% =  (-1)^{|f_1||f_2|}\left(\left[\epsilon^As_A f_2(\bx)\right]f_1(\bx) + f_2(\bx)\left[\epsilon^As_A f_1(\bx)\right]\right)\\[8pt]
% = f_1(\bx-\ba_A)\left[\epsilon^As_A f_2(\bx)\right]
%+ \left[\epsilon^As_A f_1(\bx)\right]f_2(\bx+\ba_A),
%\end{array}
%\end{equation}
\begin{eqnarray}
&&\epsilon^As_A\left[f_1(\bx)f_2(\bx)\right]  \\ %[8pt]
&&= \epsilon^As_A\left[(-1)^{|f_1||f_2|}f_2(\bx)f_1(\bx)\right]  \nonumber \\ %[8pt]
&& =  (-1)^{|f_1||f_2|}\left(\left[\epsilon^As_A f_2(\bx)\right]f_1(\bx) + f_2(\bx)\left[\epsilon^As_A f_1(\bx)\right]\right) \nonumber \\ %[8pt]
&& = f_1(\bx-\ba_A)\left[\epsilon^As_A f_2(\bx)\right] + \left[\epsilon^As_A f_1(\bx)\right]f_2(\bx+\ba_A), \nonumber
\end{eqnarray}
where the last equality follows due to the non-commutativity. Since the two
expressions for the susy transformation of $f_1f_2$ are not equal, 
the action of susy transformations on products is not well defined in the NC approach.

%%%%%%%%%%%%%%%%%%%%%%%%%%%%%%%%%%%%%%%%%%%%%%%%%%%%%%%%%%%
\section{The Link Approach}
\subsection{General Ideas}

Inspired by the NC approach the same
authors have proposed a novel
link formulation of
twisted lattice supersymmetry in
\cite{Dadda3}. This construction again has the merit of
appearing to retain {\it all} continuum supersymmetries on the
lattice.
Instead of thinking in terms of non-commuting supercoordinates $x^{\mu}$
and $\theta^A$, the $\theta^A$ are given a link variable interpretation. 
They now describe constant link variables associated to the links
$(x,x+a_A)$, $\theta^A$  being their common value:
\begin{equation}
     \epsfig{file=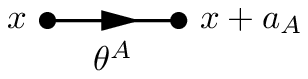, height=3.25ex} 
\end{equation}
Objects on the right of $\theta_A$ are now forced to be at $x+a_A$, objects on the left at $x$,
thereby {\it replacing} the non-commutativity relation 
\be
[\bx,\theta^{A}] = \ba_A\theta^{A} \Leftrightarrow \bx\theta^A = \theta^A(\bx+\ba_A) \quad \mbox{(no sum)}.
\ee
In a similar way the $\frac{\partial}{\partial \theta^A}$ are associated
to constant link variables and can be viewed as the same link
variable as $\theta^A$, but with opposite link orientation.
The standard algebraic properties of the $\theta^A$ and $\frac{\partial}
{\partial \theta^A}$ variables are now interpreted as link relations,
for instance $ \theta^A \theta^B = -\theta^B \theta^A$ reads
\be
\theta^A_{x+a_A+a_B,x+a_B} \theta^B_{x+a_B,x} =  -\theta^B_{x+a_B+a_A,x+a_A}\theta^A_{x+a_A,x}.
\label{theta}
\ee

Also the difference operator $\Delta_{\pm\mu}$ gets the interpretation of a constant link variable,
on the link $(x+n_{\mu},x)$. It thus follows that also the supercharges have a link interpretation.
For example, in case of  the twisted ${\cal N}=D=2$ algebra, the supercharge $Q$ reads
\be
Q_{x,x-a}=\frac{\partial}{\partial \theta}_{x,x-a} + \frac{i}{2} (\theta^\mu)_{x,x+a_\mu} (\Delta_{+\mu})_{x+a_\mu,x-a}, 
\ee
and the algebra itself reads
\be
&&Q_{x+n_{\mu},x+a_{\mu}}(Q_{\mu})_{x+a_{\mu},x} + (Q_{\mu})_{x+n_{\mu},x+a}Q_{x+a,x} \nonumber\\
&&\quad \quad\quad \quad\quad \quad = i(\Delta_{\mu})_{x+n_{\mu},x},\nonumber\\
&&\tilde{Q}_{x,x+\tilde{a}}(Q_{\mu})_{x+\tilde{a},x+\tilde{a}+a_{\mu}} + (Q_{\mu})_{x,x+a_{\mu}}\tilde{Q}_{x+a_{\mu},x+a_{\mu}+\tilde{a}} \nonumber\\
&&\quad\quad\quad \quad\quad \quad= -i\epsilon_{\mu\nu}(\Delta_{-\nu})_{x-n_{\nu},x}.
\ee
The consistency conditions for the link approach are the same as for the NC approach, namely
equation (\ref{consiscond}).

Unfortunately, a full consistent set of calculation rules for the link approach has never been presented. 
For instance, it is not clear how to consistently multiply two superfields.
This should be
contrasted with the non-commutative constructions where the calculation rules are the usual ones plus
a non-commutativity between the bosonic and fermionic coordinates of superspace. 
%It goes of course without saying that this also means that the link and non-commutativity
%approaches are two not fully equivalent approaches.

However, working on a component field level we will nevertheless
show that it is possible
to define a Leibniz rule for taking susy variations of products of fields which
allows one to construct lattice actions which are formally invariant under
the twisted lattice supersymmetries.
The supersymmetry transformations $s_A$ are treated as link variables, like the supercharges $Q_A$.
All component fields $\varphi_{x,x+a_\varphi}$
are treated as link fields, albeit including the degenerate 
case $a_\varphi=0$ in which case we speak of a site field.
A supersymmetry transformation will thus map a site or a link field onto another site or link field in such a way that 
the supersymmetry algebra is preserved. 
The lattice action is then built out of the site and link component fields.
 
Since the link approach incorporates link fields naturally 
into the discretization of a theory, it forms 
an excellent playground for lattice gauge theory where gauge fields $A_{\mu}$ 
are put on the lattice as link fields $U_{n_{\mu}} \simeq e^{iA_{\mu}}$.
 
In \cite{Dadda3} the lattice formulation of twisted ${\cal N}=D=2$ Super Yang-Mills theory is treated precisely along the lines described above. This is discussed in more detail in the next section.

%%%%%%%%%%

\subsection{Twisted ${\cal N}=D=2$ SYM on the Lattice}

In this section the discretization of twisted ${\cal N}=D=2$ SYM following the 
link approach is summarized.

To gauge the twisted ${\cal N}=D=2$ lattice theory introduced before, the constant link variables
$\Delta_{\pm}$ and $Q_A$ are replaced 
with corresponding gauge degrees of freedom\footnote{The conventions used in this paper differ from now on in some cases from the conventions used in \cite{Dadda3}. This is because two different conventions are used in that paper, whereas here we stick to one.}:
\be
(\Delta_{\pm \mu})_{x\pm n_{\mu},x}  \rightarrow  \mp (\mathcal{U}_{\pm \mu})_{x,x\pm n_{\mu}}, \\
 (Q_A)_{x,x-a_A} \rightarrow (\nabla_A)_{x,x+a_A}.
\label{corresp}
\ee
The link fields 
$(\mathcal{U}_{\pm \mu})$ and $\nabla_A$ are $x$ 
(or better link) dependent elements of the gauge group, 
just like gauge links in ordinary lattice gauge theory\footnote{Note however
that the links $\U_{\pm\mu}$ are to be thought of as exponentials of a complex vector
potential whose imaginary part will give rise to the usual scalar fields of
extended supersymmetry in the continuum limit.}.
The gauge transformation of these link variables are  given by 
\begin{eqnarray}
(\U_{\pm\mu})_{x,x\pm n_{\mu}} &\rightarrow&
G_{x}(\U_{\pm\mu})_{x,x\pm n_{\mu}}G^{-1}_{x\pm n_{\mu}},\\
(\D_{A})_{x,x+a_A} &\rightarrow& G_{x}(\D_{A})_{x,x+a_{A}}G^{-1}_{x+a_A},
\end{eqnarray}
where $G_{x}$ denotes the finite gauge transformation at the site $x$.

The following twisted ${\cal N}=D=2$ susy constraints follow from `gauging' the susy algebra
 (\ref{discrsusy}),
\begin{eqnarray}
\{\D,\D_{\mu}\}_{x,x+a+a_{\mu}} &=& + i (\U_{+\mu})_{x,x+n_{\mu}}, 
\label{gauge1} \\[2pt]
\{\tD,\D_{\mu}\}_{x,x+\tilde{a}+a_{\mu}} 
&=& -i\epsilon_{\mu\nu}\ (\U_{-\nu})_{x,x-n_{\nu}}, \label{gauge2}
\end{eqnarray}
where the left hand side of (\ref{gauge1})--(\ref{gauge2}) should be 
understood as link  anti-commutators, as in (\ref{theta}).
The equations (\ref{consiscond}) are obviously crucial for consistency.

Playing with Jacobi identities of the link matrices one can see that 
one may consistently define the following 
non-vanishing fermionic link fields:
\begin{eqnarray}
[\D_{\mu},\U_{+\nu}]_{x,x+a_{\mu}+n_{\nu}} 
&\equiv& -\epsilon_{\mu\nu}(\tilde{\rho})_{x,x-\tilde{a}}, \label{lf1}\\ [2pt]
[\D_{\mu},\U_{-\nu}]_{x,x+a_{\mu}-n_{\nu}} 
&\equiv& -\delta_{\mu\nu}(\rho)_{x,x-a}, \label{lf2} \\ [2pt]
\epsilon_{\mu\nu}[\D,\U_{-\nu}]_{x,x+a-n_{\nu}} 
&\equiv& -\epsilon_{\mu\nu}(\lambda_{\nu})_{x,x-a_{\nu}}\label{lf3} \\ [2pt]
&=& -[\tD,\U_{+\mu}]_{x,x+\tilde{a}+n_{\mu}}, \label{lf4}
\end{eqnarray}
which are then ${\cal N}=2$ twisted fermions. 
The full twisted ${\cal N}=2$ multiplet on the lattice is given by
\be
(\U_{\pm}, \rho, \tilde{\rho}, \lambda_{\mu}, K),
\ee
where 
\be
K_{x,x} \equiv \frac{1}{2}\{ \nabla_{\mu} , \lambda_{\mu} \}_{x,x}
\ee 
is an auxiliary (site) field, needed
for the supersymmetry algebra to close off-shell. 
By construction all the link and site fields in the multiplet are elements of the gauge group
and transform in the same way under gauge transformations as the 
$\U_{\pm}$ and $\nabla_A$. 
The fields $\U_{\pm}$ and $K$ are bosonic, the others are fermionic.
\\
\newline
Supersymmetry transformations of link fields are defined by
\begin{eqnarray}
(s_{A}\varphi)_{x,x+a_{\varphi}+a_A} = 
s_{A}(\varphi)_{x,x+a_{\varphi}}\equiv
[\D_{A},\varphi\}_{x,x+a_{\varphi}+a_A},
\label{susytrafo}
\end{eqnarray}
where $(\varphi)_{x,x+a_{\varphi}}$ denotes one of the component fields in 
$(\U_{\pm\mu},\rho,\tilde{\rho},\lambda_{\mu},K)$, or a product of them.
This can be worked out further to
\be
(s_A \varphi)_{x,x+a_{\varphi}+a_A} & = &(\nabla_A)_{x,a+a_A}\varphi_{x+a_A,x+a_A+a_{\varphi}} \\ \nonumber
&&-(-)^{|\varphi|} \varphi_{x,x+a_{\varphi}}(\nabla_A)_{x+a_{\varphi},x+a_{\varphi}+a_A}, 
\ee
and represented pictorially as
\be
     \epsfig{file=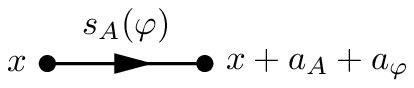, height=3.25ex} & = &
     \raisebox{-1.5ex}{\epsfig{file=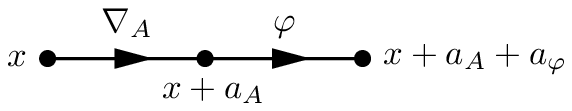, height=4.25ex}}  \\ \nonumber
  && - (-)^{|\varphi|} \,\raisebox{-1.5ex}{\epsfig{file=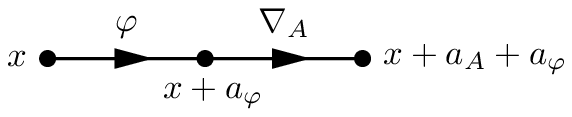, height=4.25ex}}.
\ee
 
For doing calculations one needs to know how a product of the (matrix valued) link fields transforms in terms
of the transformations of single fields, in other words, one needs to know a Leibniz rule.
It can easily be derived how something like  
\begin{equation}
     \epsfig{file=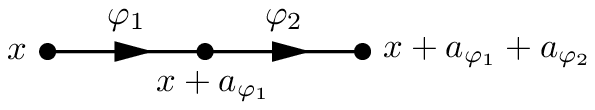, height=4.25ex}
\end{equation}
transforms, where the $\varphi_i$ are single link fields. Using the definition (\ref{susytrafo}), and by writing things 
out explicitly and adding and subtracting at the same time
\be
(-1)^{|\varphi_1|}\,\raisebox{-1.5ex}{\epsfig{file=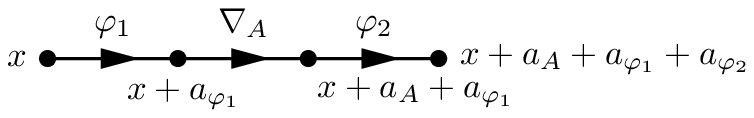, height=4.25ex}}, 
\ee
it follows that 
%The Leibniz rule should of course be of the well known general form
%\be
%s_A(\varphi_1\varphi_2) = (s_A\varphi_1)\varphi_2 +(-1)^{|\varphi_1|}\varphi_1(s_A\varphi_2).
%\ee
%Taking into account the link nature of the fields and the supersymmetry transformation, one is lead to the following rule
\be
&&(s_A \varphi_1\varphi_2)_{x,x+a_A+a_{\varphi_1}+a_{\varphi_2}} \label{guessedLrule}\\
&& =   (s_A\varphi_1)_{x,x+a_A+a_{\varphi_1}}(\varphi_2)_{x+a_A+a_{\varphi_1},x+a_A+a_{\varphi_1}+a_{\varphi_2}} \nonumber\\ %[10pt]
&&\quad +(-1)^{|\varphi_1|}(\varphi_1)_{x,x+a_{\varphi_1}}(s_A\varphi_2)_{x+a_{\varphi_1},x+a_A+a_{\varphi_1}+a_{\varphi_2}}. \nonumber
\ee
This is a Leibniz rule of the well known general form, 
but  as in the NC construction,
modified by shifts in the arguments of the fields - in this case the start and end points of link fields.
Pictorially it can be expressed as follows:
\be
&&s_A(\raisebox{-1.5ex}{\epsfig{file=phi1phi2-b.eps, height=3.5ex}}) \\
&& \quad = \raisebox{-1.5ex}{\epsfig{file=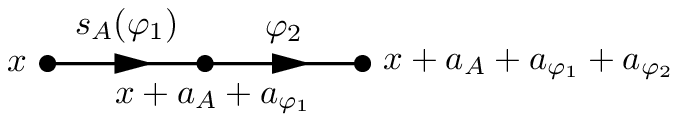, height=4.25ex}}\nonumber\\ \nonumber %[10pt]
&&\quad\quad+(-1)^{|\varphi_1|}\,\raisebox{-1.5ex}{\epsfig{file=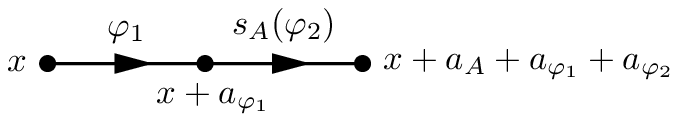, height=4.25ex}}.
\ee
\\
\newline
Using the definition of the supersymmetry transformations (\ref{susytrafo}) and the twisted ${\cal N}=D=2$ constraints
(\ref{gauge1}-\ref{gauge2}), one can deduce the susy transformations of the different component fields
explicitly by using Jacobi identities. The result is shown in table \ref{susytrafo-s} for the transformation $s$. 
\begin{table}[t]
\begin{tabular}{|c||c|}
\hline
& $s$ \\ \hline
$\U_{+\nu}$ & $0$ \\
$\U_{-\nu}$ & $-\lambda_{\nu}$  \\ 
$\lambda_{\nu}$ & $0$ \\
$\rho$ & $-\frac{i}{2}[\U_{+\rho},\U_{-\rho}]-K$  \\
$\tilde{\rho}$ & $-\frac{i}{2}\epsilon_{\rho\sigma}[\U_{+\rho},\U_{+\sigma}]$  \\
$K$ & $+\frac{i}{2}[\U_{+\rho},\lambda_{\rho}]$ \\ \hline
\end{tabular}
\caption{SUSY transformation of ${\cal N}=2$ lattice SYM multiplet.}
\label{susytrafo-s}
\end{table}
The full result can be found in \cite{Dadda3}.
Arguing in a similar way, it can be seen 
that the following supersymmetry algebra must hold on the lattice
\begin{eqnarray}
\{s,s_{\mu}\}(\varphi)_{x,x+a_{\varphi}} 
&=& +i[\U_{+\mu},\varphi]_{x,x+a_{\varphi}+n_{\mu}},
\label{uptogage1} \\[2pt]
\{\tilde{s},s_{\mu}\}(\varphi)_{x,x+a_{\varphi}} &=& -i\epsilon_{\mu\nu}
[\U_{-\nu},\varphi]_{x,x+a_{\varphi}-n_{\nu}},
\label{uptogage2} \\[2pt]
s^{2}(\varphi)_{x,x+a_{\varphi}}&=&
\tilde{s}^{2}(\varphi)_{x,x+a_{\varphi}} = 0,\\
\{s,\tilde{s}\}(\varphi)_{x,x+a_{\varphi}}\ 
&=& \{s_{\mu},s_{\nu}\}(\varphi)_{x,x+a_{\varphi}}\ =\ 0,
\end{eqnarray}
where $\varphi$ denotes any (product of) component(s) of the multiplet
$(\U_{\pm\mu},\rho,\tilde{\rho},\lambda_{\mu},K)$.
%\begin{comment}%%%
%Using the definition of the supersymmetry transformations (\ref{susytrafo}) and the twisted $N=D=2$ constraints
%(\ref{gauge1}-\ref{gauge2}), one can deduce that the following supersymmetry algebra must hold on the lattice
%\begin{eqnarray}
%\{s,s_{\mu}\}(\varphi)_{x+a_{\varphi},x} 
%&=& +i[\U_{+\mu},\varphi]_{x+a_{\varphi}+n_{\mu},x},
%\label{uptogage1} \\[2pt]
%\{\tilde{s},s_{\mu}\}(\varphi)_{x+a_{\varphi},x} &=& +i\epsilon_{\mu\nu}
%[\U_{-\nu},\varphi]_{x+a_{\varphi}-n_{\nu},x},
%\label{uptogage2} \\[2pt]
%s^{2}(\varphi)_{x+a_{\varphi},x}&=&
%\tilde{s}^{2}(\varphi)_{x+a_{\varphi},x} = 0,\\
%\{s,\tilde{s}\}(\varphi)_{x+a_{\varphi},x}\ 
%&=& \{s_{\mu},s_{\nu}\}(\varphi)_{x+a_{\varphi},x}\ =\ 0,
%\end{eqnarray}
%(link indices wrong!)
%where $\varphi$ denotes any (product of) component(s) of the multiplet
%$(\U_{\pm\mu},\rho,\tilde{\rho},\lambda_{\mu},K)$.
%
%This algebra is derived by arguments like 
%\be
%0 & = & [\nabla,[\nabla,\varphi\}\} +  [\nabla,[\nabla,\varphi\}\} +[\{\nabla,\nabla\},\varphi\}  \\
 %  & = & [\nabla,[\nabla,\varphi\}\} +  [\nabla,[\nabla,\varphi\}\} \\
 %  & \equiv & 2 s^2(\varphi),
%\ee
%where the link nature of the objects has been suppressed. 

%Arguing in a similar way, also the susy transformations of the different component fields
%can be derived explicitly. The result is shown in Table ....
%\end{comment}%%%
Finally the action of twisted ${\cal N}=D=2$ SYM on the lattice is given by
\be
S &\equiv& \frac{1}{4}\sum_{x}\mathrm{Tr}\ 
s\tilde{s}\epsilon_{\mu\nu}s_{\mu}s_{\nu}\ \U_{+\mu}\U_{-\mu},
\label{action}
\ee
where the summation over $x$ should also cover
the additional lattice sites introduced by the shift parameters $a_A$.
For the symmetric choice of these parameters (Figure~\ref{symmetric_a}) this means a sum over the integer 
sites  $(m_1,m_2)$ and over the half integer sites $(m_1+\frac{1}{2},m_2+\frac{1}{2})$.

The fact that the action is given by the consecutive action of all the susy transformations on 
$\U_{+\mu}\U_{-\mu}$ suggests that the action is by construction
invariant under all the supersymmetries due to the nature of the lattice susy algebra. 
Furthermore, the closed loop nature of $\U_{+\mu}\U_{-\mu}$ plus the fact that the total shift
of $s\tilde{s}\epsilon_{\mu\nu}s_{\mu}s_{\nu}$ is given by $a+\tilde{a}+a_1+a_2 = 0$, makes sure
that the action consists of a sum over closed loops, 
and is therefore manifestly gauge invariant.

%\begin{comment}
%However, if one wants to check the susy invariance of the action, or if one
%wants to deduce the explicit form of the action (\ref{a4}) from (\ref{a1}), 
%one needs to know how the susy transformations $s_A$ act
%on products of links. 
%In other words, one needs to know a Leibniz rule for the susy transformations.
%Unfortunately, it is not clear what this Leibniz rule is. Following the same definition 
%(\ref{susytrafo}) for the transformation of products of link fields as for link component fields 
%themselves, a Leibniz rule can easily be deduced. However, this Leibniz rule is not
%well defined when it comes to products of link fields that form a closed loop.
%In the next section these problems are discussed in detail.
%\end{comment}

%%%%%%%%%%%%%%%%%%%%%%%%%%%%%%%%%%%%%%%%%%%%%%%%%%%%%%%%

\section{The Inconsistency}
\label{inconsistency}

It can easily be seen from an expression like (\ref{guessedLrule}) that the Leibniz rule is not invariant
under flipping the order of $\varphi_1$ and $\varphi_2$, due to the shifts in the arguments of the 
fields just as in the NC approach. 
However due to the link nature of the fields,
an expression like 
$(\varphi_1)_{x,x+a_{\varphi_1}}(\varphi_2)_{x+a_{\varphi_1},x+a_{\varphi_1}+a_{\varphi_2}}$
can not be exchanged for 
$(\varphi_2)_{x+a_{\varphi_1},x+a_{\varphi_1}+a_{\varphi_2}}(\varphi_1)_{x,x+a_{\varphi_1}}$,
since the latter expression is not well defined and in general
the two matrices $\varphi_1$ and $\varphi_2$ do not commute.
Thus, in contrast to the
NC formulation the Leibniz rule (\ref{guessedLrule}) seems to be well defined in this link approach.

However, it is not. Consider the case of a closed loop,  $a_{\varphi_1}=-a_{\varphi_2}$.
Both $(\varphi_1)_{x,x+a_{\varphi_1}}(\varphi_2)_{x+a_{\varphi_1},x}$
and $(\varphi_2)_{x+a_{\varphi_1},x}(\varphi_1)_{x,x+a_{\varphi_1}}$ are well defined and are  graphically represented by the same loop diagram
\be
\raisebox{-1ex}{\epsfig{file=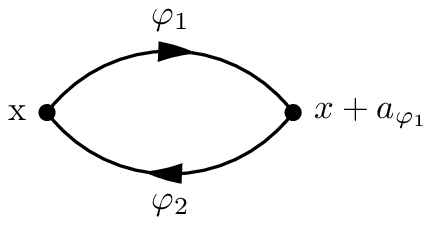, height=7.25ex}}.
\label{lus}
\ee
Of particular interest are traces over closed loops since
they occur in the action. Upon taking
the trace these two expressions are equal up to a factor $(-1)^{|\varphi_1||\varphi_2|}$.

Consider now the susy transformation of the first ordering of the
fields. We find
\be
&&s_A\mathrm{Tr}[(\varphi_1)_{x,x+a_{\varphi_1}}(\varphi_2)_{x+a_{\varphi_1},x}] \label{sphi1phi2}\\[8pt]
&&=  \mathrm{Tr}[(s_A\varphi_1)_{x,x+a_A+a_{\varphi_1}}(\varphi_2)_{x+a_A+a_{\varphi_1},x+a_A}]\nonumber\\[8pt]
&&\quad \quad +(-1)^{|\varphi_1|} \mathrm{Tr}[(\varphi_1)_{x,x+a_{\varphi_1}}(s_A\varphi_2)_{x+a_{\varphi_1},x+a_A}] \nonumber\\[8pt]
&&= \mathrm{Tr}\,\raisebox{-1,5ex}{\epsfig{file=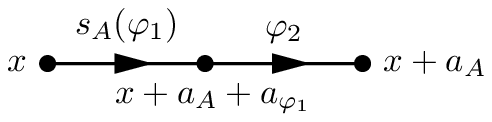, height=4.25ex}}\nonumber\\[8pt]
&&\quad\quad+(-)^{|\varphi_1|}\, \mathrm{Tr}\,\raisebox{-1,5ex}{\epsfig{file=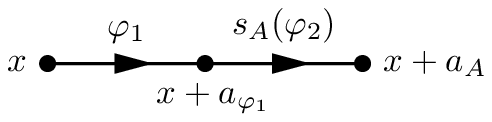, height=4.25ex}}. \nonumber
\ee
But susy transforming the second ordering leads to the expression (up to the  factor 
$(-1)^{|\varphi_1||\varphi_2|}$)
\be
&&s_A \mathrm{Tr}[(\varphi_2)_{x+a_{\varphi_1},x}(\varphi_1)_{x,x+a_{\varphi_1}}]\label{sphi2phi1}\\[8pt]
&&=   \mathrm{Tr}[(s_A\varphi_2)_{x+a_{\varphi_1},x+a_A}(\varphi_1)_{x+a_A, x+a_A+a_{\varphi_1}}]\nonumber\\[8pt]
&&\quad\quad+(-1)^{|\varphi_2|} \mathrm{Tr}[(\varphi_2)_{x+a_{\varphi_1},x}(s_A\varphi_1)_{x,x+a_{\varphi_1}+a_A}] \nonumber\\[8pt]
&&= \mathrm{Tr}\,\raisebox{-1,5ex}{\epsfig{file=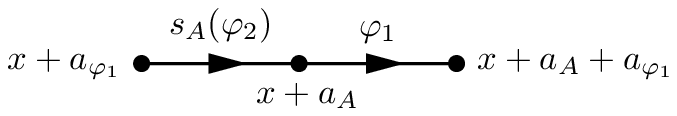, height=4.25ex}}\nonumber\\[8pt]
&&\quad\quad+(-)^{|\varphi_2|}\, \mathrm{Tr}\,\raisebox{-1,3ex}{\epsfig{file=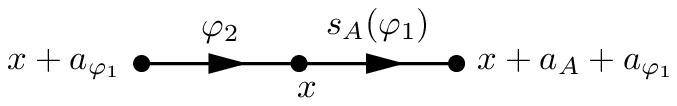, height=4.25ex}}.\nonumber
\ee
These expressions are clearly not identical. 
They do not consist of the same fields.
In fact the first ordering leads
to a field living on the link $(x,x+a_A)$ while the second gives a field
residing on the link $(x+a_{\varphi_1},x+a_{\varphi_1}+a_A)$.
Therefore the supersymmetric transformation of such a 
gauge invariant loop is not
well defined.

The technical reason for the inconsistency comes from the link nature:
when transforming the first field in a product like (\ref{sphi1phi2}) or (\ref{sphi2phi1}), the
second field has to be shifted by the shift of $s_A$.
Hence the supersymmetry transformations treat fields in products according
to their order and reversing the order then leads to the shown contradiction.
It is easy to see that this inconsistency applies to closed loop products 
with two or more fields (i.e.\ closed loops with two or more legs).

A special case of the situation discussed above deserves a little more attention: 
$a_{\varphi_1}=-a_{\varphi_2} = 0$.
We are now considering two matrix valued fields $F$ and $G$ that live on lattice sites.
Following the Leibniz rule, the susy transformation of the product $F_{x,x}G_{x,x}$ is given by
\be
s_A \mathrm{Tr}[F_{x,x}G_{x,x}] & =  &  \mathrm{Tr}[(s_AF)_{x,x+a_A}G_{x+a_A,x+a_A}]  \label{sFG}\\[8pt]\nonumber
&& \quad\quad + (-1)^{|F|} \mathrm{Tr}[F_{x,x}(s_AG)_{x,x+a_A}].
\ee
But since we trace over $F$ and $G$, we can  
flip the order of the fields in the product, and the following is obtained (up to the factor
$(-1)^{|F||G|}$)
\be
s_A \mathrm{Tr}[G_{x,x}F_{x,x}] & = &  \mathrm{Tr}[(s_AG)_{x,x+a_A}F_{x+a_A,x+a_A}]  \label{sGF}\\[8pt]\nonumber
&&\quad \quad + (-1)^{|G|} \mathrm{Tr}[G_{x,x}(s_AF)_{x,x+a_A}].
\ee
This is almost exactly the inconsistency as it was 
encountered in the NC approach,
see \cite{FalkMe}.
This inconsistency affects all the supercharges
associated with the symmetric solution
for the shift parameters (\ref{symmsolution}).

Notice, however that there is no inconsistency in the case
when the shift parameter $a_A=0$. This occurs for example 
for one of the supersymmetries of the
solution given by equation (\ref{azerosolution}).
In this case the susy 
transformation $s$ maps a closed loop 
to a sum of closed loops, and it is easy to see using cyclic
permutation of the trace that all orderings of the fields yield exactly
the same expression for the susy variation. Such a supercharge has a site
character and is hence similar to the conserved supercharges 
which appear in both
\cite{Kaplan} and \cite{Catterall}. As in
that work, a maximum of one supercharge
can be implemented exactly in this lattice model. Furthermore such a charge
behaves as a scalar under Lorentz transformations.
This can be done for any of the four supersymmetries by working with a
suitable choice of the shift parameters. The other three supersymmetries
will have non zero shift parameter, and will suffer from the inconsistency.
\\
\newline
Of course an interesting question is how this inconsistency affects the 
supersymmetric transformation of the action.
If one derives the action from its definition (\ref{action})
by ignoring the inconsistency explained above, i.e.\ by just
taking the fields in the order in which they come, not changing the order in a product using
the trace cyclicity, the following form of the action is obtained (taking into account the equations (\ref{consiscond}))
\begin{eqnarray}
S & = &\sum_{x}\mathrm{Tr}
\biggl[ \\
&&( \frac{i}{2}[\U_{+\mu},\U_{-\mu}]-K )_{x,x} (- \frac{i}{2}[\U_{+\nu},\U_{-\nu}] -K )_{x,x} 
\nonumber\\%[2pt]
&& -\frac{1}{4}\epsilon_{\mu\nu}\epsilon_{\rho\sigma}
[\U_{+\mu},\U_{+\nu}]_{x,x+n_{\mu}+n_{\nu}}
[\U_{-\rho},\U_{-\sigma}]_{x+n_{\rho}+n_{\sigma},x}
\nonumber\\%[2pt]
&& -i[\U_{+\mu},\lambda_{\mu}]_{x,x+a}(\rho)_{x+a,x} + \nonumber\\%[2pt]
&& - i(\tilde{\rho})_{x,x-\tilde{a}}\epsilon_{\mu\nu}
[\U_{-\mu},\lambda_{\nu}]_{x-\tilde{a},x}.\nonumber
\biggr].
\end{eqnarray}
Susy transforming this form of the action under $s$ gives the expected result, zero.
However, using the the cyclicity of the trace to write the action in the form given in \cite{Dadda3},
\begin{eqnarray}
S & = &\sum_{x}\mathrm{Tr}
\biggl[  \\
&&\frac{1}{4}[\U_{+\mu},\U_{-\mu}]_{x,x}[\U_{+\nu},\U_{-\nu}]_{x,x} +K_{x,x}^{2} \nonumber\\%[2pt]
&& -\frac{1}{4}\epsilon_{\mu\nu}\epsilon_{\rho\sigma}[\U_{+\mu},\U_{+\nu}]_{x,x+n_{\mu}+n_{\nu}}
[\U_{-\rho},\U_{-\sigma}]_{x+n_{\rho}+n_{\sigma},x}
 \nonumber\\%[2pt]
&&  -i[\U_{+\mu},\lambda_{\mu}]_{x,x+a}(\rho)_{x+a,x}  + \nonumber\\%[2pt]
&& - i(\tilde{\rho})_{x,x-\tilde{a}}\epsilon_{\mu\nu}
[\U_{-\mu},\lambda_{\nu}]_{x-\tilde{a},x}\nonumber
\biggr],
\end{eqnarray}
and susy transforming this form of the action under $s$, it leads to the following 
\be s S & = & \frac{1}{2}\sum_{x}\mathrm{Tr} \biggl[\frac{1}{2}[\U_{+\mu},\lambda_{\mu}]_{x,x+a}[\U_{+\nu},\U_{-\nu}]_{x+a,x+a} 
\label{sSnonzero} \\
&&-\frac{1}{2}[\U_{+\mu},\U_{-\mu}]_{x, x}[\U_{+\nu},\lambda_{\nu}]_{x,x+a} + \nonumber\\ 
&& iK_{x,x}[\U_{+\mu},\lambda_{\mu}]_{x,x+a} -i[\U_{+\mu},\lambda_{\mu}]_{x,x+a}K_{x+a,x+a}
\biggr].\nonumber
\label{ssnotzero}
\ee
In the case $a \neq 0$, the situation here is 
thus very much like that encountered in the NC approach 
\cite{FalkMe}, where
the action written without 
interchanging fields in a product was naively invariant, 
but no longer upon changing the order. 
In the case $a = 0$, the expression (\ref{sSnonzero}) does reduce to zero, 
and thus, in this situation,
there is again no inconsistency associated
with the supersymmetry $s$.
However, the other supersymmetries will now have non-zero shift
parameters and variation with respect to them will be ill-defined.
\\
\newline
For illustration, consider one term in the action in more detail.
The following is part of the action 
\be
&&-i\sum_x \mathrm{Tr}\left[(\tilde{\rho})_{x+\tilde{a},x}(\U_{-1})_{x,x-n_1}(\lambda_{2})_{x-n_1,x+\tilde{a}} \right]\nonumber\\
&&\quad\quad=-i\sum_x \mathrm{Tr}\raisebox{-4,5ex}{\epsfig{file=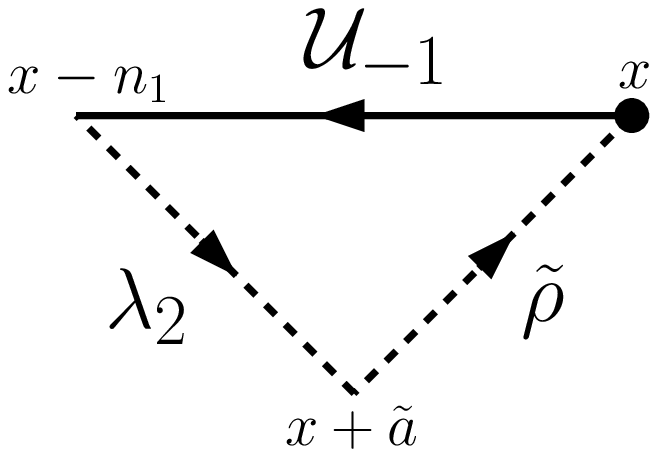, height=10.25ex}}
\label{looppartaction}
\ee
Transforming the above loop as it stands under $s$ leads to
\be
&&i\sum_x \mathrm{Tr}\left[ (\tilde{\rho})_{x+\tilde{a},x}(\lambda_1)_{x,x-a_1} (\lambda_2)_{x-a_1,x+a+\tilde{a}} \quad +   \right. \label{sloop1} \\
&&\left. \frac{i}{2}\epsilon_{\rho\sigma}[\U_{+\rho},\U_{+\sigma}]_{x+\tilde{a},x+a}(\U_{-1})_{x+a,x-a_1} 
(\lambda_{2})_{x-a_1,x+a+\tilde{a}} \right], \nonumber
\ee
which is a link from $x+\tilde{a}$ to $x+a+\tilde{a}$.
Using the trace cyclicity, (\ref{looppartaction}) can equivalently be written as
\be
-i\sum_x \mathrm{Tr}\left[(\U_{-1})_{x,x-n_1}(\lambda_{2})_{x-n_1,x+\tilde{a}}(\tilde{\rho})_{x+\tilde{a},x}\right].
\label{looppartaction-traced}
\ee
Transforming this expression leads to
\be
&&i\sum_x \mathrm{Tr}\left[ (\lambda_1)_{x,x-a_1} (\lambda_2)_{x-a_1,x+a+\tilde{a}}(\tilde{\rho})_{x+a+\tilde{a},x+a}   \quad
\right. \label{sloop2}\\
&&\left. +(\U_{-1})_{x,x-n_1} (\lambda_{2})_{x-n_1,x+\tilde{a}}(\frac{i}{2})\epsilon_{\rho\sigma}[\U_{+\rho},\U_{+\sigma}]_{x+\tilde{a},x+a}  \right], \nonumber
\ee
which is a link from $x$ to $x+a$. 
Choosing $a = 0$, the expressions (\ref{sloop1}) and (\ref{sloop2}) 
are two equivalent ways of 
writing the same loop, and there is no inconsistency for $s$. 
Choosing $a \neq 0$, the two expressions are different links, 
and the inconsistency is present. 

%%%%%%%%%%%%%%%%%%%%%%%%%%%%%%%%%%%%%%%%%%%%%%%%%%

\section{Discussion and Conclusion}

In a series of recent papers D'Adda et al.\ \cite{Dadda, Dadda3} have 
developed novel approaches
to discretizing certain supersymmetric theories with the goal of
preserving {\it all} continuum (twisted) supersymmetries. 

Two
approaches have been constructed both of which modify the
supersymmetry variation of products of fields so as to make it compatible
with a modified lattice Leibniz rule. In the approach described in \cite{Dadda}
this property is ensured by introducing a non-commutativity between
superspace coordinates. An inconsistency in this approach was pointed out
in \cite{FalkMe} and we have summarized this problem again here.

This paper focuses on an analysis of the second approach, termed
the link construction, in which the fields live either on links or on sites of a lattice and
transform under supersymmetry variation into fields of opposite
Grassmann character living on
neighboring links or sites. The supersymmetry transformations also either have a link or a site
character.
The precise link or site nature of the fields and the supersymmetry transformations depends on the
choice of shift parameters, and ensures that a closed lattice supersymmetry algebra can be constructed
incorporating lattice difference operators \cite{Dadda3}. In this paper we summarize
this construction and point to another
apparent inconsistency of this
approach which is similar to that encountered with the non-commutative
construction. While the arguments are quite general we illustrate
them by referring to the lattice action for ${\cal N}=2$ super Yang-Mills
theory described in \cite{Dadda3}. 

The problem is best exposed by considering the supersymmetry variation
of a gauge invariant Wilson loop. If a supersymmetry variation has a link character, 
we show that the result of
supersymmetry variation of this closed loop is not invariant under an initial
cyclic permutation of the fields in the loop.
Since the action is built out of traces over such loops, 
this inconsistency is visible at the level
of the action -- different cyclic permutations of terms in the
action yield the same action written in different ways whose supersymmetry variations 
are not equal: they differ from zero to non-zero. 

In the ${\cal N} = D = 2$ model the shift parameters are constrained such that
at most one of them is zero. 
Choosing one of the parameters zero, the corresponding susy variation
has a site character and can be placed on the lattice consistently. 
The other three have a link character and their action on gauge invariant
loops is not well defined. 
Choosing none of them zero, all will suffer from this inconsistency.

Furthermore, 
if a susy transformation has a link character, 
the variation of the action under this transformation 
will always be a link variable.  
Since link variables are not gauge invariant, 
this means that the expectation value of
the supersymmetry variation
of the action will be zero, regardless of the way the action is written.
This, in turn, means that 
the partition function will actually be susy invariant.
At first glance this reasoning appears
to be a way out of the inconsistency, but it is not. 
Indeed, this observation
actually implies that the susy variation of {\it any}  gauge invariant
lattice action will have vanishing expectation
value -- clearly a non-physical result.

As a final remark, the transformations described here
and in \cite{Dadda3} do not contain an infinitessimal Grassman parameter - it is
possible that the introduction of such a parameter with link character
may alleviate some of the problems \cite{last}.

%%%%%%%%%%%%%%%%%%%%%%%%%%%%%%%%%%%%%%%%%
 
\section*{Acknowledgements}

We are grateful to Anosh Joseph and Jan-Willem van Holten for useful discussions. 
FB was supported by DFG (FOR 465), SC in part by DOE grant DE-FG02-85ER40237, and 
MK by FOM.

%%%%%%%%%%%%%%%%%%%%%%%%

{}

\end{document}